\documentclass[12pt]{article}
\usepackage{amsmath}
\usepackage{amssymb}
\usepackage{bm}
\usepackage[dvipdfmx]{graphicx,color,hyperref}

\catcode `@=11 \@addtoreset{equation}{section} \catcode `@=12

\begin{document}
\setlength{\oddsidemargin}{0cm}
\setlength{\baselineskip}{7mm}

\begin{titlepage}

\begin{center}
{\LARGE
 Bound on Lyapunov exponent  in $c=1$ matrix model
  }
\end{center}
\vspace{0.2cm}
\baselineskip 18pt 
\renewcommand{\thefootnote}{\fnsymbol{footnote}}

\begin{center}
Takeshi {\sc Morita}\footnote{%
E-mail address: morita.takeshi@shizuoka.ac.jp
} 

\renewcommand{\thefootnote}{\arabic{footnote}}
\setcounter{footnote}{0}

\vspace{0.4cm}

{\small\it
Department of Physics,
Shizuoka University, \\
836 Ohya, Suruga-ku, Shizuoka 422-8529, Japan}\\

\vspace{0.4cm}

{\small\it
   Graduate School of Science and Technology, Shizuoka University\\
836 Ohya, Suruga-ku, Shizuoka 422-8529, Japan
}

\end{center}


\vspace{1.5cm}

\begin{abstract}

Classical particle motions in an inverse harmonic potential show the exponential sensitivity to  initial conditions,  where the Lyapunov exponent $\lambda_L$ is uniquely fixed by the shape of the potential.
Hence, if we naively apply the bound on the Lyapunov exponent $\lambda_L \le 2\pi T/ \hbar$ to this system, 
it predicts the existence of the bound on temperature (the lowest temperature) $T \ge \hbar \lambda_L/  2\pi$ and the system cannot be taken to be zero temperature when $\hbar \neq 0$.
This seems a puzzle because particle motions in an inverse harmonic potential should be realized without introducing any temperature but this inequality does not allow it.
In this article, we study this problem in $N$ non-relativistic free fermions in an inverse harmonic potential ($c=1$ matrix model).
We find that thermal radiation is {\em induced} when we consider the system in a semi-classical regime even though the system is not thermal at the classical level.
This is analogous to the thermal radiation of black holes, which are classically non-thermal but behave as thermal baths quantum mechanically.
We also show that the temperature of the radiation in our model saturates the inequality, and thus, the system saturates the bound on the Lyapunov exponent, although the system is free and integrable.
Besides, this radiation is related to acoustic Hawking radiation of the fermi fluid.

\end{abstract}


\end{titlepage}

\section{Introduction}

Understanding quantum gravity is the one of the most important problem in theoretical physics.
Through the developments in the gauge/gravity correspondence \cite{Maldacena:1997re, Itzhaki:1998dd}, many people expect that large-$N$ gauge theories may illuminate natures of quantum gravity.
However only special classes of large-$N$ gauge theories which possess certain properties may describe the gravity, and understanding what are essential properties in gauge theories to have their gravity duals is a crucial question.

Recently the idea of the maximal Lyapunov exponent was proposed, and it might capture the one of the essence of the gauge/gravity correspondence \cite{Maldacena:2015waa}.
The authors of \cite{Maldacena:2015waa} conjectured that thermal many-body quantum systems have an upper bound on the Lyapunov exponent: 
\begin{align}
\lambda_L \le \frac{2\pi T}{\hbar}  ,
\label{L-bound}
\end{align}
 where $T$ is temperature of the systems and $\lambda_L$ is the Lyapunov exponent. (We take $k_B=1$ in this article.)
Particularly, if a field theory at a finite temperature has the dual black hole geometry, the gravity calculation predicts that the field theory should saturate this bound $\lambda_L = 2\pi T/ \hbar$ \cite{Shenker:2013pqa, Shenker:2013yza}.
(This would be related to the conjecture that the black hole may provide the fastest scrambler in nature \cite{Sekino:2008he}.)
This is called maximal Lyapunov exponent, and the properties of this bound is actively being studied.
One remarkable example is the SYK model \cite{Sachdev:1992fk, Kitaev:2015}.
We can show that this model saturates the bound from the field theory calculation, and now people are exploring what the dual gravity of this model is \cite{Almheiri:2014cka,Maldacena:2016hyu,Jensen:2016pah,Maldacena:2016upp,Engelsoy:2016xyb,Cvetic:2016eiv,Blake:2016jnn,Mandal:2017thl,Stanford:2017thb,Mezei:2017kmw,Forste:2017kwy,Das:2017pif,Taylor:2017dly,Mertens:2017mtv,Grumiller:2017qao,Sarosi:2017ykf,Das:2017hrt, Das:2017wae, Haehl:2017pak,Forste:2017apw}.

The purpose of this article is to have a deeper understanding of the bound \eqref{L-bound}.
Many previous works on the bound \eqref{L-bound} studied finite temperature systems.
However, in the conventional studies of chaos in dynamical systems, people mainly considered deterministic dynamics, for example, driven pendulum motions, and thus the systems are at zero temperature.
Obviously, the inequality \eqref{L-bound} at zero temperature seems problematic, because the right hand side becomes zero and the inequality requires the Lyapunov exponent to be zero, while the Lyapunov exponent can be non-zero in chaotic systems even at zero temperature.
One answer for this puzzle is saying that the bound  \eqref{L-bound} should be restricted to quantum many body systems at finite temperature only.
Indeed, this is the setup in which the bound \eqref{L-bound} was proved in \cite{Maldacena:2015waa}, and we cannot apply the proof to such conventional chaotic systems at  zero temperature.
Nevertheless, it may be interesting to ask the question of whether the bound \eqref{L-bound} plays any role in the conventional cases.

In this article, we study this problem in one-dimensional point particle motion in an inverse harmonic potential
\begin{align}
m \ddot x(t) = -V'(x), \qquad V(x)=-\frac{\alpha }{2}x^2.
\label{inverse-EOM}
\end{align}
Classical solution of this equation is given by
\begin{align}
x(t)=c_1 e^{\sqrt{\alpha/m}t} + c_2 e^{-\sqrt{\alpha/m}t},
\label{particle-chaos}
\end{align}
where $c_1$ and $c_2$ are constants determined by initial conditions.
Thus this solution shows the exponential sensitivity to initial conditions with Lyapunov exponent\footnote{
	In classical mechanics, the Lyapunov exponent is defined by
	\begin{align}
		\delta q(t) \sim \delta q(0) \exp(\lambda_L t) .
		\label{chaos}
	\end{align}
	Here $q(t)$ is the value of an observable at time $t$ and $\delta q(t)$ is the deviation of  $q(t)$  through the change of the initial condition by $\delta q(0)$.
In the recent studies of the chaos bound \cite{Maldacena:2015waa,Kurchan:2016nju,Tsuji:2017fxs,Scaffidi:2017ghs,Liu:2018}, the Lyapunov exponents defined through the out-of-time-ordered correlator (OTOC) \cite{1969JETP...28.1200L} are mainly investigated.
The connection between the Lyapunov exponent defined by OTOC and the classical Lyapunov exponent \eqref{chaos} are discussed in \cite{Kitaev:2015-2, Maldacena:2015waa}.
We assume that these two exponents agree up to $O(\hbar)$ corrections in the semi-classical regime.
},
\begin{align}
\lambda_L=\sqrt{\frac{\alpha}{m}}.
\label{Lyapunov}
\end{align}
Of course the motion (\ref{particle-chaos}) is too simple and not even chaotic.
However, through this simple model, we may gain some insights into the bound (\ref{L-bound}) in  actual chaotic systems.

Now let us consider the inequality \eqref{L-bound} in our model.
Since our model has the finite Lyapunov exponent \eqref{Lyapunov} at the classical level, it may be better to rewrite the bound \eqref{L-bound} as  \cite{Kurchan:2016nju}\footnote{Here we have assumed a semi-classical approximation, and the quantum correction to the Lyapunov exponent \eqref{Lyapunov} is small and can be ignored.
We will discuss the validity of this approximation at the end of  section \ref{sec-c=1}.
}
\begin{align}
T \ge \frac{\hbar}{2\pi} \lambda_L =  \frac{\hbar}{2\pi} \sqrt{\frac{\alpha}{m}}.
\label{T-bound}
\end{align}
This relation predicts the existence of the lower bound on temperature in this system,
\begin{align}
T_L := \frac{\hbar}{2\pi} \lambda_L =\frac{\hbar}{2\pi} \sqrt{\frac{\alpha}{m}}.
\label{T-mini}
\end{align}
 In the strict classical limit $\hbar \to 0$, this bound does not play any role, since the right hand side becomes 0.
 Obviously, the classical solution \eqref{particle-chaos} is nothing to do with thermodynamics and it is consistent with the inequality \eqref{T-bound} with $T=0$.
 

 However,  the relation \eqref{T-bound} becomes non-trivial in a semi-classical regime, in which the right hand side becomes non-zero.
We will argue that indeed the lowest temperature $T_L$ plays an important role in our model.
To see it clearly, we consider $N$ non-relativistic free fermions in the inverse harmonic potential (\ref{inverse-EOM}).
This model is related to a one-dimensional matrix quantum mechanics called $c=1$ matrix model which describes a two dimensional gravity through the non-critical string theory (see, e.g. the review
of \cite{Klebanov:1991qa,Ginsparg:1993is,Polchinski:1994mb}).

\begin{figure}
\begin{center}
\includegraphics{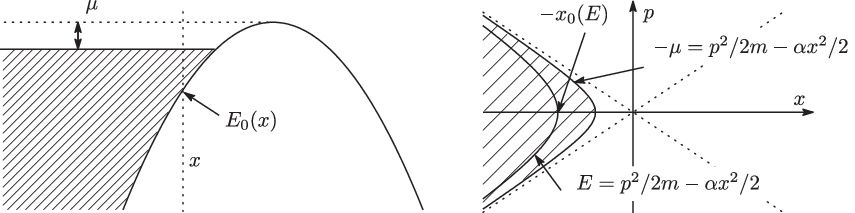}
\caption{The sketch of the fermions in the inverse harmonic potential which corresponds to the bosonic non-critical string theory. $-\mu$ is the fermi energy.
(RIGHT) The configuration of the fermions in the phase space.
The region occupied by the fermions is called ``droplet". 
}
\label{Fig-string} 
\end{center}
\end{figure}

We consider the configuration of the fermions which describes the bosonic non-critical string theory (figure \ref{Fig-string}).
(In the following arguments, the correspondence to the string theory is not important.)
Although this configuration is stable in the strict classical limit, it is unstable via the quantum tunneling of the fermions. 
We will show that this instability {\it induces} a thermal flux with the temperature $T_L$ (\ref{T-mini}) in the asymptotic region $x \to - \infty$ which flows from around $x=0$.
This radiation is an analogous of the Hawking radiation \cite{Hawking:1974sw, Hawking:1974rv} in the sense that the system is not thermal classically whereas the thermal spectrum arises quantum mechanically.
Therefore the inverse harmonic potential can be regarded as a black body with the temperature $T_L$ for the observer in the asymptotic region, and the temperature bound (\ref{T-bound}) may be saturated.

We will also argue that this radiation is related to acoustic Hawking radiation in a supersonic fluid \cite{Unruh:1980cg}.
(A related study in the context of condensed matter physics has been done in \cite{Giovanazzi:2004zv} and developed in \cite{Giovanazzi:2006nd,Parola:2017qpn}.)
The collective motion of the $N$ fermions are described as a fermi fluid, and it classically obeys the hydrodynamic equations \cite{Dhar:1992rs,Dhar:1992hr,Mandal:2013id}.
Hence, if the fluid velocity exceeds the speed of sound, acoustic Hawking radiation may be emitted.
We will argue how the supersonic region can be realized in the $c=1$ matrix model, and show that the acoustic event horizon arises only at the tip of the inverse harmonic potential $x=0$ if the flow is stationary.
Thus the radiation may arise from $x=0$ when we turn on the quantum corrections.
We will show that the temperature of this acoustic Hawking radiation is always given by the lowest temperature $T_L$, and argue the connection to the instability of the bosonic non-critical string theory.

In such a way, the bound on the Lyapunov exponent \eqref{L-bound}, which is equivalent to the temperature bound \eqref{T-bound}, plays an interesting role even in our simple inverse harmonic potential model \eqref{inverse-EOM}, which is not chaotic and not thermal at the classical level, through the Hawking radiation.
Hence we expect that the bound \eqref{L-bound} may be relevant in the conventional chaotic systems, which are deterministic and not thermal\footnote{After the author submitted this manuscript on arxiv, he developed the idea of this manuscript and wrote another paper \cite{Morita:2019bfr} in which the role of the inequality \eqref{L-bound} in the conventional chaotic systems was revealed. }.

Note that related works on the particle creations in the $c=1$ matrix model have been done in the context of the non-critical string theory \cite{Takayanagi:2003sm,Douglas:2003up}.
They include time dependent background and tachyon condensation \cite{Karczmarek:2004ph,Das:2004hw,Mukhopadhyay:2004ff,Karczmarek:2004yc}, and high energy scattering and black hole physics \cite{Martinec:2004qt, Friess:2004tq,Karczmarek:2004bw, Banks:2015mya, Betzios:2016yaq}.

The organization of this article is as follows.
In section \ref{sec-c=1}, we argue the appearance of the thermal radiation with the temperature $T_L$ (\ref{T-mini}) in the $c=1$ matrix model in the configuration of the fermions corresponding to the bosonic non-critical string theory.
In section \ref{sec-acoustic-HR}, we show the connection between this radiation and acoustic Hawking radiation. 
A brief review on the derivation of the acoustic Hawking radiation is included.
Section \ref{sec-discussion} is discussions and future problems.


\section{The $c=1$ matrix model and the temperature bound }
\label{sec-c=1}

We consider the $c=1$ matrix model which is equivalent to $N$ non-relativistic free fermions.
The fermions classically obey the equation of motion (\ref{inverse-EOM}), and the 
Hamiltonian for the single fermion is given by 
\begin{align}
\hat H =  -\frac{\hbar^2}{2m} \frac{\partial^2}{\partial x^2} - \frac{\alpha}{2} x^2.
\label{H-c=1}
\end{align}
This Hamiltonian is unbounded from below, and usually we introduce suitable boundaries  so that the fermions are confined.
These boundaries are irrelevant as far as we focus on phenomena around $x=0$.
In this section, we introduce infinite potential walls at $x=\pm L$.

We consider the configuration of the fermions which describes the two dimensional bosonic non-critical string theory.
We put all the fermions on the left of the inverse harmonic potential, and tune the parameters $N,\alpha,m$ and $L$ so that $N \to \infty$ and the fermi energy $-\mu$ ($\mu>0$) is close but below zero.
We sketch this configuration in figure \ref{Fig-string}.

At the classical level, this configuration is zero temperature because there is no thermal fluctuations.
However, as we will see soon, once we turn on quantum effects, quantum fluctuations mimic {\it thermal fluctuations} and radiation that has a thermal spectrum with  the temperature $T_L$ defined in (\ref{T-mini}) is induced.

Let us derive this thermal radiation.
Obviously the above configuration cannot be stationary quantum mechanically. 
The fermions will leak to the right side of the potential through the quantum tunneling.
Such a tunneling effect is significant for the fermions close to the fermi surface.
For a single fermion with energy $E$ $(E \le - \mu)$, the tunneling probability $P_T(E)$  can be calculated as \cite{Moore:1991zv},
\begin{align}
 \quad P_T(E)= \frac{1}{\exp\left( -\frac{2\pi}{\hbar} \sqrt{\frac{m}{\alpha}} E \right)+1}.
\label{PT}
\end{align}
The derivation is shown in appendix \ref{app-QM}.
This formula is suggestive.
If we regard $-E(>0)$ as energy, $P_T(E)$ is identical to the Fermi-Dirac distribution with the  temperature
\begin{align}
T = \frac{\hbar}{2\pi}\sqrt{\frac{\alpha}{m}},
\label{T-mini-2}
\end{align}
which precisely agrees with the lowest temperature $T_L$ (\ref{T-mini})\footnote{This tunneling probability is for a single particle, and the appearance of the Fermi-Dirac distribution is nothing to do with the fact that we are considering the fermions.}. 
Thus the tunneling of the fermions might cause thermal radiation.

In order to confirm this thermal property, we evaluate the contributions of the tunneling to physical quantities and ask whether they are really thermal or not.
We will investigate the local density of an observable $O(\hat{x},\hat{p})$.
First we will compute this quantity in the classical limit ($\hbar \to 0$).
Then we will calculate the same quantity including the quantum effect and, by comparing these two results, we will evaluate the contributions of the tunneling.

Let us start from the calculation in the classical limit $\hbar \to 0$.
In this limit, the fermion with negative energy $E$ is confined in the left region $-L \le x \le - x_0(E)$,
where $x_0(E):= \sqrt{-2E/\alpha} $ is the classical turning point. (See figure \ref{Fig-string}.)
The WKB wave function for such a fermion is given by
\begin{align}
\phi_E(x)= \frac{C(E)}{\sqrt{p(E,x)}} \frac{1}{\sqrt{2}} \left( e^{\frac{i}{\hbar} \int_{-L}^x dy \, p(E,y) }-e^{-\frac{i}{\hbar} \int_{-L}^x dy \, p(E,y) } \right),
\label{WKB-wave}
\end{align}
where $p(E,x)$ is the classical momentum for the particle,
\begin{align}
p(E,x)=\sqrt{2m\left(E+\frac{\alpha}{2}x^2\right)},
\label{moment}
\end{align}
and $C(E)$ is the normalization factor satisfying
\begin{align}
|C(E)|^2 = \left( \int_{-L}^{-x_0(E)} \frac{dx}{ |p(E,x)| } \right)^{-1} ,
\end{align}
so that $\phi_E(x)$ is normalized to 1.
Through the boundary condition $\phi_E(-L)=\phi_E(-x_0(E))=0$, energy is quantized as
\begin{align}
\int_{-L}^{-x_0(E)} dx\,| p(E,x)| = \pi  \hbar n, \qquad (n=1,2,\cdots),
\label{BS-cond}
\end{align}
where we have ignored the fragment terms.
By using this WKB wave function, we evaluate the expectation value  of the observable $O(\hat{x},\hat{p})$ for the $N$ fermions as
\begin{align}
\langle O(\hat{x},\hat{p}) \rangle_N =& \sum_{i=1}^N \langle i| O(\hat{x},\hat{p}) |i \rangle= \int_{E_G}^{-\mu} dE \, \rho(E)  \int_{-L}^{-x_0(E)} dx \, |\phi_E (x)|^2
O(x,p(E,x)),
\label{VEV-O-all} 
\end{align}
where $E_G:=- \alpha L^2/2 $ is the ground state energy and $\rho(E)$ is the energy density given by
\begin{align}
\rho(E) = \frac{\partial n}{\partial E}= \frac{m}{\pi \hbar } \int_{-L}^{-x_0(E)} \frac{dx}{| p(E,x)| },
\label{E-density}
\end{align}
through the quantization condition (\ref{BS-cond}).
Here we have assumed that $N$ is sufficiently large and the spectrum is continuous.
From (\ref{VEV-O-all}), the density of the observable $O(\hat{x},\hat{p})$ at position $x$ is given by
\begin{align}
 \int_{E_0(x)}^{-\mu} dE \, \rho(E)   |\phi_E (x)|^2
O(x,p(E,x))=
 \frac{m}{\pi \hbar}
 \int_{E_0(x)}^{-\mu} \frac{d E}{|p(E,x)|} \, O(x,p(E,x)).
 \label{O-classical}
\end{align}
Here $E_0(x):=-\alpha x^2/2$ is the minimum energy at given $x$. (See figure \ref{Fig-string}.)

Now we evaluate the quantum correction to the density (\ref{O-classical}).
In the WKB wave function (\ref{WKB-wave}), 
$\exp\left(\frac{i}{h} \int^x dy \, p(y)\right)$ can be regarded as the incoming wave toward the inverse harmonic potential, while $\exp\left(-\frac{i}{h} \int^x dy \, p(y)\right)$  can be regarded as the reflected wave.
Then through the tunneling effect (\ref{PT}), a part of the incoming wave will penetrate to the right side of the potential, and
the reflected wave will be decreased by the factor $\sqrt{1-P_T(E)}$.
Hence the WKB wave function (\ref{WKB-wave}) will be modified as
\begin{align}
\phi_E(x) \to \frac{C(E)}{\sqrt{p(E,x)}} \frac{1}{\sqrt{2}} \left( e^{\frac{i}{\hbar} \int_{-L}^x dy \, p(E,y) }- \sqrt{1-P_T(E)} e^{-\frac{i}{\hbar} \int_{-L}^x dy \, p(E,y) } \right), \quad (-L \ll x \ll -x_0(E)).
\end{align}
Note that this result is not reliable near the turning point $ -x_0(E)$, since the WKB approximation does not work.
Also we have assumed that the boundary $x=-L$ is sufficiently far, and we can ignore the further reflections there. 
This correction modifies the left hand side of (\ref{O-classical}), and the density of the observable $O(\hat{x},\hat{p})$ becomes,
\begin{align}
 \frac{m}{\pi \hbar}
 \int_{E_0(x)}^{-\mu}  \frac{dE}{|p(E,x)|}  \, 
\left(1- \frac{1}{2} P_T(E)\right) O(x,p(E,x)).
 \end{align}
By subtracting the classical result (\ref{O-classical}) from this result, 
we obtain the contributions of the quantum tunneling to the density  of the observable $O(\hat{x},\hat{p})$ as
\begin{align}
- \frac{m}{2\pi \hbar} \int_{E_0(x)}^{-\mu} \frac{dE}{|p(E,x)|}  \,   \frac{ O(x,p(E,x))}{e^{ -\beta_L E }+1} = - \frac{m}{2\pi \hbar} \int_\mu^\infty   \frac{d E}{|p(-E,x)|}  \frac{ O(x,p(-E,x))}{e^{ \beta_L E }+1},
 \label{VEV-O-semi}
\end{align}
where $\beta_L$ denotes the inverse temperature $1/T_L$ and we have approximated $E_0(x) \approx  -\infty$ by assuming that we are considering the density in a far region\footnote{In this region, the tunneling of the particle with the energy $E_0(x)$ is sufficiently suppressed.} $-L \ll  x \ll  - (\hbar^2/m \alpha)^{1/4}$.
We also flipped $E \to - E$ in the last step.

We can interpret the result (\ref{VEV-O-semi}) as follows. 
Due to the tunneling, the holes with positive energy $E$ $(E \ge \mu)$ appear in the fermi sea.
They move to the left and contribute to the quantity $O(x,p)$.
Importantly, (\ref{VEV-O-semi}) shows that these holes are thermally excited and obey the Fermi-Dirac distribution at the temperature $T_L$.
Therefore the quantum tunneling indeed induces thermal radiation.

For example, if we choose $O(\hat{x},\hat{p})=\hat H$ in (\ref{VEV-O-semi}), we obtain the energy density at $x$ in the far region
\begin{align}
- \frac{m}{2\pi \hbar} \int_0^\infty   \frac{d E}{|p(-E,x)|}
 \frac{-E }{e^{ \beta_L E }+1} = \frac{1}{ (- x)} \frac{T_L}{48}\left(1 + O\left(\frac{T_L}{\alpha x^2}\right) \right).
\end{align}
Here we have taken $\mu=0$ for simplicity.
The obtained energy density is positive reflecting the fact that the holes carry the positive energies.

Our result indicates that the inverse harmonic potential plays a role of a black body at the temperature $T_L$.
This result reminds us the Hawking radiation \cite{Hawking:1974sw, Hawking:1974rv} in which the thermal excitation arises through the quantum effect even though the system is not thermal in the classical limit.
Indeed we will see that the thermal excitation in the matrix model is related to the acoustic Hawking radiation in quantum fluid.

Notice that, since the thermal radiation at the temperature $T_L$ arises, 
 the temperature bound (\ref{T-bound}) may be saturated in this system\footnote{
 	We should emphasize that this temperature is obtained through the probability distribution of the observables (\ref{VEV-O-semi}), and it is not a genuine temperature of this system.
 	Indeed, since the $c=1$ matrix model is free and integrable, genuine thermalization does not occur.
 	(In integrable systems, thermalization to the Generalized Gibbs ensemble (GGE) \cite{Mandal:2013id, PhysRevLett.98.050405,  KMM} may occur. 
 	It may be interesting to consider the role of the temperature $T_L$ in the context of GGE.)
 	In this sense, the saturation of the bound \eqref{L-bound} in our model might not be directly related to the one studied in \cite{Maldacena:2015waa}. On the other hand, since our temperature is related to Hawking radiation, it may capture an essence of the bound.}. 
(The scattering of the particle in the inverse harmonic potential would occur within the Ehrenfest’s time\footnote{
	If we regard the  the Ehrenfest’s time as the scrambling time $t_*$ \cite{Maldacena:2015waa}, and apply the scaling relation $\hbar =1/N$ in the $c=1$ matrix model \cite{Klebanov:1991qa,Ginsparg:1993is,Polchinski:1994mb}, we obtain the scrambling time $t_* \sim (\frac{\hbar \beta_L}{2\pi})\log N $. 
Then the $c=1$ matrix model may be regarded as a fast scrambler \cite{Sekino:2008he}.
In \cite{Maldacena:2015waa}, the dissipation time $t_d \sim 1/\lambda_L $ is also discussed, and exponential developments \eqref{Lyapunov} appear in $t_d< t < t_*$ in semi-classical chaotic systems.
Particularly we expect $t_d \sim \hbar \beta$ in strongly coupled systems \cite{Maldacena:2015waa}.
Interestingly, although  the $c=1$ matrix model is free, it shows $t_d \sim 1/\lambda_L \sim \hbar \beta_L $ through \eqref{T-mini-2} and imitates the relation in strongly coupled systems.} $t \sim (1/\lambda_L)\log(1/\hbar) $. 
 Hence the Lyapunov exponent obtained in the classical motion \eqref{Lyapunov} is reliable 	\footnote{At very late times, the system will settle down into the state in which the half of the fermions are in the left side of the potential and the other half are in the right.
 	There the formula (\ref{VEV-O-semi}) would not work. 
 	Thus the argument of the temperature bound  would be valid only for the initial stage of the time evolution starting from the meta-stable configuration sketched in figure \ref{Fig-string} such that the formula (\ref{VEV-O-semi}) is reliable. (On the other hand, the Lyapunov exponent \eqref{Lyapunov} is always well defined classically, since any particle motion obeys \eqref{particle-chaos}.)
 }.)

\section{Thermal radiation in the $c=1$ matrix model and acoustic Hawking radiation}
\label{sec-acoustic-HR}

We will show that the thermal radiation in the $c=1$ matrix model argued in the previous section is related to the acoustic Hawking radiation \cite{Unruh:1980cg}.
(Several arguments in this section have been done by \cite{Giovanazzi:2004zv} in the context of condensed matter physics.)

\subsection{Acoustic Hawking radiation in the $c=1$ matrix model}
\label{sec-review-acoustic}

As is well known, the fermions in the $c=1$ matrix model compose a two dimensional ideal fluid.
The fermion particle number density $\rho(x,t)$ and velocity field $v(x,t)$ classically obey the following the continuity equation and  the Euler equation with pressure $p=\hbar^2 \pi^2 \rho^3/3 m$ \cite{Dhar:1992rs,Dhar:1992hr,Mandal:2013id},
\begin{align}
&\partial_t \rho + \partial_x(\rho v)=0, \qquad
 \partial_t v + \partial_x \left(\frac{1}{2}  v^2  +  \frac{ \hbar^2 \pi^2}{2m^2} \rho^2 + \frac{1}{m}  V(x)\right)=0.
\label{euler}
\end{align}
Here $V(x)=- \alpha x^2/2$.  (But many of the arguments in this section will work for general $V(x)$.)
We ignore the infinite potential walls at $x=\pm L$ in this section.
Hence we can apply the story of the acoustic Hawking radiation to this system.

Here we briefly show the derivation of the acoustic Hawking radiation \cite{Unruh:1980cg}.
(See e.g. a review article \cite{Visser:1997ux}.)
We consider the following small $\epsilon$ expansion
\begin{align}
\rho(x,t)=& \rho_0(x,t) + \epsilon \rho_1(x,t) + \epsilon^2 \rho_2(x,t) + \cdots, \nonumber \\
v(x,t)=& v_0 (x,t) + \epsilon v_1 (x,t) + \epsilon^2 v_2 (x,t) + \cdots.
\label{expansion}
\end{align}
Here we regard that $\rho_0$ and $v_0$ describe a background current, and $\rho_1$ and $v_1$ describe the phonons propagating on it, and $\rho_n$ and $v_n$ ($n\ge 2$) are the higher corrections.
(We have assumed that such a separation between the background and the phonons is possible.)
We will see that when we turn on the quantum effect, $\rho_1$ and $v_1$ may show a thermal excitation if there is a supersonic region in the background.

\subsubsection{Acoustic geometry for phonons}
\label{sec-acoustic-GR}

In order to derive the acoustic Hawking radiation, we consider the equations for the phonons.
By substituting the expansion (\ref{expansion}) into the hydrodynamic equation (\ref{euler}) and expanding them with respect to $\epsilon$, we will obtain the equations for $\rho_n$ and $v_n$ order by order.
Especially, at order $\epsilon$, we obtain the equations for $\rho_1$ and $v_1$,
\begin{align}
&\partial_t \rho_1 + \partial_x\left(\rho_0 v_1+v_0 \rho_1 \right)=0, \qquad
 \partial_t v_1 + \partial_x\left( v_0 v_1  +  \frac{\hbar^2 \pi^2 \rho_0}{m^2}   \rho_1 \right)=0.
\label{eq-v1}
\end{align}
Here we introduce the velocity potential $\psi$ such that $v_1=- \partial_x \psi$.
Then from the second equation, we obtain
\begin{align} 
\rho_1= \frac{m^2}{\hbar^2 \pi^2 \rho_0} \left( \partial_t \psi + v_0 \partial_x \psi \right).
\end{align}
By substituting this result to the first equation of (\ref{eq-v1}), we obtain the wave equation for the velocity potential $\psi$ as
 \begin{align}
0=& \partial_\mu \sqrt{-g} g^{\mu\nu} \partial_\nu \psi,
\label{wave-phonon}
\end{align}
where $g_{\mu\nu}$ ($\mu,\nu=t,x$) is given by
\begin{align}
g_{\mu\nu} = \frac1{ \gamma \rho_0} 
\begin{pmatrix}
c^2-v_0^2 & v_0 \\
v_0 & -1
\end{pmatrix},
\label{g_mn}
\end{align}
and $c$ denotes the speed of sound defined by
 \begin{align}
 c(x,t) :=  \sqrt{\frac{\partial p}{\partial \rho}}= \frac{ \hbar \pi }{m} \rho_0 (x,t).
 \label{sound speed}
 \end{align}
Here $p$ is the pressure defined above (\ref{euler}).
Note that $\gamma$ in $g_{\mu\nu}$ is an arbitrary constant, which cannot be fixed \footnote{In two dimension, the wave equation (\ref{wave-phonon}) is invariant under a scale transformation, and we have the ambiguity $\gamma$.}. 
This wave equation is identical to the wave equation of a scalar field on a curved manifold with the metric (\ref{g_mn}).
For this reason, $g_{\mu\nu}$ is called acoustic metric.

Importantly, if the fluid velocity $|v_0|$ is increasing along $x$-direction and exceeds the speed of sound $c$  at a certain point, 
$c=|v_0|$ occurs there and the $(t,t)$-component of the metric (\ref{g_mn}) vanishes similar to the event horizon of  a black hole.
Correspondingly, we can show that no phonon can propagate from the supersonic region $(c<|v_0|) $ to the subsonic region $(c>|v_0|)$ across the point $c=|v_0|$.
For this reason, the point $c=|v_0|$ is called acoustic event horizon. 
Then we expect that the phonon $\psi$ in the subsonic region ($c<|v_0| $) may be thermally excited quantum mechanically similar to the Hawking radiation.
We will confirm it later.

\subsubsection{Acoustic black holes and white holes}
\label{sec-BH-WH}

\begin{figure}
\begin{center}
\includegraphics{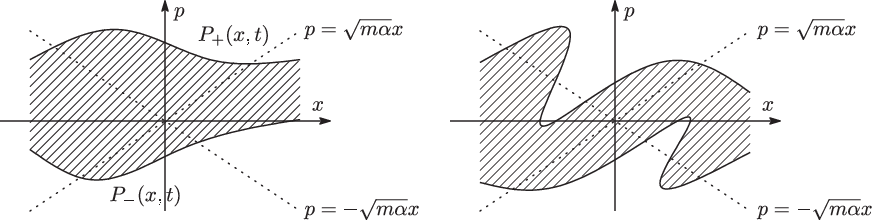}
\caption{The sketch of the fermions in the phase space.
They compose the droplet(s) in the phase space.
The doted lines describe the separatrices $p=\pm \sqrt{\alpha/m}x$. 
(LEFT)
$P_+(x,t)$ and $P_-(x,t)$ (\ref{P_+}) are the upper and lower boundaries of the droplet, respectively.
(RIGHT)
If the boundaries of the droplet have the multiple values as functions of $x$, 
$P_+(x,t)$ and $P_-(x,t)$ are not enough to define the boundaries.
Such a multiple value configuration is called ``fold" \cite{Ginsparg:1993is, Polchinski:1991uq} and 
the hydrodynamic equation (\ref{euler}) does not work there \cite{KMM, PhysRevLett.109.260602}.
We do not consider this case in this article.
}
\label{Fig-phase} 
\end{center}
\end{figure}

When does $c=|v_0|$ occur in our matrix model?
In order to see it, 
it is convenient to define $P_{\pm}$ as
\begin{align}
P_\pm(x,t):= m v(x,t) \pm \hbar \pi \rho(x,t).
\label{P_+}
\end{align}
It means that
\begin{align}
\rho(x,t) & =\frac{1}{2\pi \hbar} (P_+(x,t) - P_-(x,t)), \qquad
v(x,t)  = \frac1{2m} (P_+(x,t) + P_-(x,t)).
\label{quad-densities}
\end{align}
Here $P_{\pm}$ describe the boundaries of the droplet of the fermions in the phase space as shown in figure \ref{Fig-phase}. (We have implicitly excluded the case in which the folds \cite{Ginsparg:1993is, Polchinski:1991uq} appear on the boundaries of the droplet.)
Then from (\ref{sound speed}) and (\ref{quad-densities}), we can rewrite $c$ and $v_0$ by using $P_{0\pm}$,
\begin{align}
c\pm v_0= \pm \frac{1}{m} P_{0\pm}(x,t),
\label{c-v-P}
\end{align}
where we have expanded 
\begin{align}
P_{\pm}=P_{0\pm}+ \epsilon P_{1\pm}+ \epsilon^2 P_{2\pm}+ \cdots,
\label{expand-P}
\end{align}
corresponding to the expansion (\ref{expansion}). 
Therefore $c= |v_0|$ occurs when $P_{0+}$ or $P_{0-}$ vanishes, and
an acoustic event horizon may appear there.

Now we solve  the fluid equation (\ref{euler}), and find the acoustic metric involving an acoustic event horizon.
By substituting (\ref{quad-densities}) to (\ref{euler}), we obtain equations for $P_\pm$, 
\begin{align}
\partial_t P_+ + \partial_x\left(  \frac{1}{2m} P_+^2 + V(x) \right) =0,\; \qquad
\partial_t P_- + \partial_x\left( \frac{1}{2m} P_-^2 + V(x) \right)=0.
\label{eom-P}
\end{align}
By using the expansion (\ref{expand-P}) in these equations and considering $O(\epsilon^0)$ terms, we obtain equations for $P_{0\pm}$ which are just $P_{\pm} \to P_{0\pm}$ in (\ref{eom-P}).
We assume that the background is stationary $\partial_t P_{0\pm}=0$. 
Then the solution of these equations are given by
\begin{align}
 P_{0\pm}  = \sigma \sqrt{ 2 m ( E_\pm - V(x))},
\end{align}
where $E_{\pm}$ are constants and $\sigma$ is $+1$ or $-1$.

\begin{figure}
\begin{center}
\includegraphics{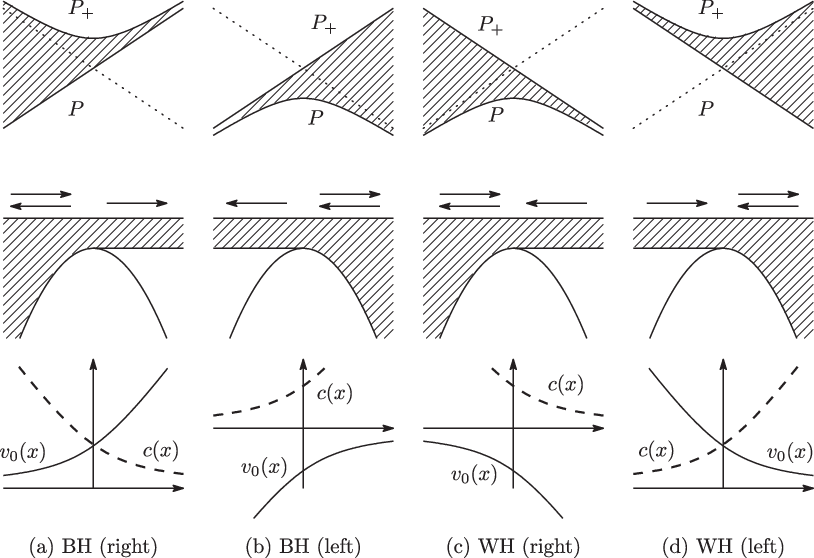}
\\
\hspace{3pt}
\caption{Four possible acoustic event horizons ($c=|v_0|$).  
(TOP) Droplets in the phase space.
(MIDDLE) The particle motion in the potential.
The arrows indicate the possible directions of the phonon propagation.
(BOTTOM) The plots of the fluid velocity $v_0$ and the speed of sound $c$ (dashed line).
(a) Black hole (right): The phonons in the supersonic region ($x>0$) cannot propagate into the subsonic region ($x<0$).
(b) Black hole (left): $x \to -x$ of the black hole (right).
(c) White hole (right): The phonons in the subsonic region ($x<0$) cannot propagate to the supersonic region ($x>0$).
(d) White hole (left): $x \to -x$ of the white hole (right).
}
\label{Fig-BH-WH} 
\end{center}
\end{figure}

To realize the acoustic event horizon, $ P_{0+} $  or  $ P_{0-} $ has to cross 0.
We can easily show that it occurs only when 
\begin{align}
 &P_{0+}  = \sqrt{ 2 m  E_+ + m \alpha x^2}, \qquad  P_{0-}  = \sigma \sqrt{ m \alpha }x, \nonumber \\
  \text{or} \qquad &P_{0+}  = \sigma \sqrt{ m \alpha }x, \qquad  P_{0-}  = - \sqrt{ 2 m  E_- + m \alpha x^2},
\end{align}
with positive $E_+$ and $E_-$.
Here the acoustic event horizon appears at $x=0$ in both cases.
Then depending on the choice of $\sigma=\pm 1$, there are four possibilities.
We call them as
 black hole (right), black hole (left), white hole (right) and  white hole (left).
See figure \ref{Fig-BH-WH}.
These names refer to the ways of the classical propagation of the phonons, and (left) and (right) are related by parity $x \to -x$.
To see the phonon propagation, we consider the black hole (right) case as an example.
As shown in figure \ref{Fig-BH-WH}, $P_{0\pm}$ in this case are given by 
\begin{align}
 P_{0+}  =  \sqrt{ 2 m  E_+ + m \alpha x^2}, \qquad  P_{0-}  =  \sqrt{m \alpha }x.
 \label{BH-right}
\end{align} 
From (\ref{c-v-P}) (and the plots of $v_0$ and $c$ in figure \ref{Fig-BH-WH}), $x<0$ is the subsonic region ($c>|v_0| $) while $x>0$ is the supersonic region ($c<|v_0| $).
Then we expect that no phonon can propagate into $x<0$ from $x>0$. Let us confirm it.
The propagation of the phonons can be read off from the wave equation (\ref{wave-phonon}).
By using new coordinates (called time-of-flight coordinates)
\begin{align}
z_{\pm}:= t - m \int^x \frac{dx'}{P_{0 \pm}(x')},
\label{TOF-coordinate}
\end{align}
the wave equation (\ref{wave-phonon}) becomes $\partial_{z_+} \partial_{z_-} \psi =0$.
Thus the solution is given by
\begin{align}
\psi=\psi_+(z_+)+\psi_-(z_-).
\label{sol-psi}
\end{align}
Then $P_{1\pm}$ become
\begin{align}
P_{1\pm} =m v_1 \pm \hbar \pi \rho_1 = \frac{2m^2}{P_{0\pm}} \psi'_\pm (z_\pm) .
\label{sol-P}
\end{align}
This relation shows (the well-known result for experts) that  the fluctuation of the boundaries of the droplet $P_{\pm}$ describes
 the excitations of the phonons and they propagate depending on $z_\pm$.
 
\begin{figure}
\begin{center}
\input{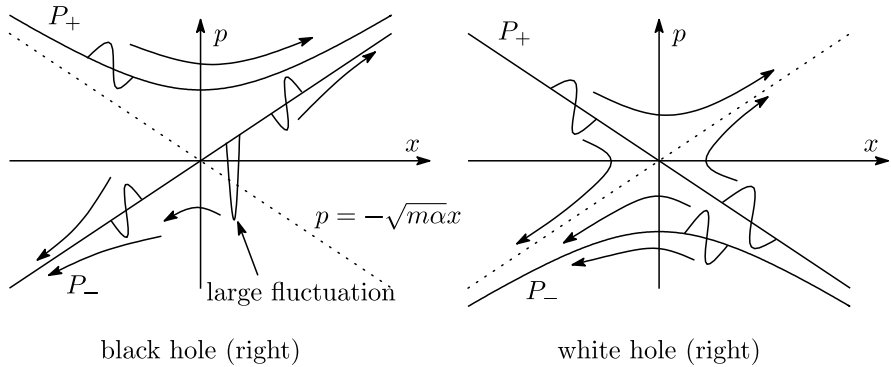}
\caption{The sketch of the propagation of finite fluctuations of $P_{\pm}$ on the black hole (right) and the white hole (right) in the phase space.
We can read off the propagation from the classical particle trajectories  $E=p^2/2m - \alpha x^2/2$ in the phase space. (Recall that the droplet is the collection of the fermionic particles, and each point in the droplet moves according to (\ref{particle-chaos}).)
In the black hole case, small fluctuations in the supersonic region ($x>0$) cannot propagate into the subsonic region ($x<0$). 
However, large fluctuations which cross the separatrix $p=-\sqrt{m\alpha}x$ can propagate.
In the white hole case, even small fluctuations on $P_+$ in the subsonic region ($x<0$) can propagate into the supersonic region ($x>0$).
It means that the ``causality" of the acoustic white hole is unstable against perturbations on $P_+$.
}
\label{Fig-fluct} 
\end{center}
\end{figure}

 In the black hole (right) case (\ref{BH-right}),  $P_{1+}(z_+)$ always propagates toward the right direction, since $P_{0+}$ is always positive in (\ref{TOF-coordinate}).
 On the other hand, we need to take care $P_{1-}(z_-)$. Since $P_{0-}$ becomes zero at $x=0$ and $z_-$ diverges there, $z_-$ covers only the left region ($x<0$) or the right region ($x>0$) as
\begin{align}
z_{-}=
\begin{cases}
  t - \sqrt{\frac{m}{\alpha}} \log x + \text{const.} & (x>0)  \\
  t - \sqrt{\frac{m}{\alpha}} \log (-x) + \text{const.} & (x<0)
\end{cases}
.
\label{tortoise}
\end{align} 
Thus $P_{1-}(z_-)$ propagates to left in $x<0$ and to right in $x>0$.
Therefore both $P_{1\pm}$ cannot propagate into $x<0$ from $x>0$ as we expected.
This is similar to the black hole geometry in which no mode can propagate to the outside from the inside.
For this reason we call this case as ``black hole (right)", by identifying the supersonic region ($x>0$) as the inside of the black hole\footnote{Note that this result is valid only for sufficiently small fluctuations.
 The ways of the propagation of the finite fluctuations of $P_{\pm}$ can  be read off from the particle trajectories as shown in figure \ref{Fig-fluct}.
Particularly if the fluctuations are large and they cross the separatrix $p=-\sqrt{m\alpha}x$, even the fluctuations in the supersonic region can propagate into the subsonic region.
Besides, it is known that arbitrary non-stationary fluctuations develop to folds at late times, and the above arguments should be modified once folds appear. }
.

Lastly, we take a detour and discuss the white hole (right) case.
In figure \ref{Fig-BH-WH} (c),  $P_{0\pm}$ is given by
\begin{align}
P_{0+}  = - \sqrt{ m \alpha }x, \qquad  P_{0-}  = - \sqrt{ 2 m  E_- + m \alpha x^2}.
\end{align}
In this case, from (\ref{c-v-P}), $v_0$ is always negative and the background current flows from the right to left.
Importantly the fluid velocity $|v_0|$ decelerates along the flow, and the supersonic flow in $x>0$ becomes subsonic at $x=0$.
As a result, $\psi$ in the subsonic region ($x<0$) cannot propagate into the supersonic region.
This is an analogue of a white hole in fluid mechanics.

However, this conclusion is valid only for $\epsilon \rightarrow 0$ limit.
 $P_{0+}$ is unstable against the perturbations and even tiny fluctuations in the subsonic region can propagate into the supersonic region as shown in figure \ref{Fig-fluct}.

\subsubsection{Acoustic Hawking radiation}
\label{sec-HR}

Now we are ready to derive the acoustic Hawking radiation.
We consider the black hole (right) case.
In this case, the classical solution of $\psi$ is given by (\ref{sol-psi}), and importantly the coordinate $z_-$ (\ref{tortoise}) covers only either $x<0$ or $x>0$. 
This is an analogue of the tortoise coordinates in the black hole geometries, and $z_-$ is not suitable to define the vacuum of the quantized phonon.

A coordinate which is well defined on  $-\infty < x < \infty$ is a Kruskal-like coordinate $U:=-e^{-\kappa z_-}=xe^{-\sqrt{\frac{\alpha}{m}}t} $, where we have defined ``the surface gravity" \footnote{
Up to an overall factor, the line element of the acoustic metric (\ref{g_mn}) can be written as
$ds^2 = c^2 f dt^2+ 2v_0 dt dx -dx^2$, where $f:=1-v_0^2/c^2= - P_{0+}P_{0-}/m^2c^2$.
Then the surface gravity $\kappa$ can be written as the familiar formula $\kappa= \frac{c}{2} \partial_x f |_{x=0}$. 
Also $\kappa$ is written as $\kappa=\frac{1}{m} \partial_x P_{0-}|_{x=0}$.
The equations in this footnote work even for a general potential $V(x)$ if the local maximum is at $x=0$ and the acoustic event horizon appears there.
},
\begin{align}
\kappa:=\sqrt{\frac{\alpha}{m}}.
\label{kappa}
\end{align}
The vacuum of the phonon $\psi_-(z_-)$ should be defined by using this coordinate.
It implies that $\psi_-$ in the asymptotic region (large negative $x$)  will be thermally excited at temperature
\begin{align}
T=\frac{\hbar \kappa}{2\pi} = \frac{\hbar }{2\pi} \sqrt{\frac{\alpha}{m}},
\label{Hawking-temp}
\end{align}
through the Bogoliubov transformation.
Here the obtained temperature precisely agrees with the lowest temperature $T_L$ (\ref{T-mini}).

For example, we can calculate the expectation value of the energy momentum tensor of the phonon by using the two dimensional conformal field theory technique \cite{Birrell:1982ix, Thorlacius:1994ip,Iso:2007kt},
\begin{align}
\langle  T_{z_- z_-}  \rangle =-\frac{c_{\psi}}{24\pi} \{ U, z_- \}=
 \frac{\hbar}{48\pi} \kappa^2, \qquad
\langle  T_{z_+ z_+}  \rangle =
 0,
  \label{EM-tensor}
\end{align}
where $\{ U, z_- \}:=U'''/U' - \frac{3}{2}(U''/U')^2  $ is the Schwarzian derivative and $c_{\psi}$ is the central charge which we have taken $c_{\psi}=1$ since phonon is a boson.
Note that $\langle  T_{z_+ z_+}  \rangle = 0$, since $z_+$ is well defined on $-\infty < x < \infty$. 

\subsection{Acoustic Hawking radiation from quantum mechanics}
\label{sec-HR-QM}

So far we have seen that the acoustic Hawking radiation occurs in the $c=1$ matrix model, since this model can be regarded as a two dimensional fluid.
Interestingly the Hawking temperature (\ref{Hawking-temp}) is given by the temperature $T_L$ which is also coincident with the temperature (\ref{T-mini-2}) derived from the quantum mechanics of the fermions in section \ref{sec-c=1}.
Although the configuration of the fermions for the bosonic non-critical string case sketched in figure \ref{Fig-string} and the acoustic Hawking radiation case sketched in figure \ref{Fig-BH-WH} are different\footnote{In terms of the fluid variables, the configuration of the bosonic non-critical string is given by $P_{0\pm}= \pm \sqrt{-2m\mu +m \alpha x^2} $, ($x<0$). See figure \ref{Fig-string} (RIGHT).}, the same temperature is obtained.
We argue why these two temperatures agree.
 
\begin{figure}
\begin{center}
\includegraphics{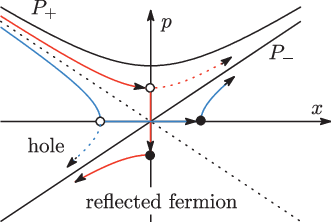}
\caption{The mechanism of the appearance of the Hawking radiation in the acoustic black hole (right) in the phase space.
The holes and the reflected fermions which are not allowed in the classical motions cause the acoustic Hawking radiation when we consider the quantum corrections.
}
\label{Fig-HR} 
\end{center}
\end{figure}

In the classical particle picture, both the bosonic non-critical string and the acoustic black hole are described by the flow of  the fermions from the left.
The difference of these two configurations are just the maximum energies.
The energy of the fermions are filled up to the fermi energy $-\mu <0$ in the bosonic non-critical string case while up to $E_+ >0$ in the acoustic black hole case.
Correspondingly all the fermions are reflected by the potential in the classical limit $\hbar \to 0$ in the bosonic non-critical string case, and, in the acoustic black hole case, they are reflected if $E<0$ and go through the potential if  $0<E \le E_+$.
As we have seen in section \ref{sec-c=1}, for the fermions with $E<0$, the holes appear in the fermi sea through the quantum tunneling, and they cause the thermal radiation.
In the acoustic black hole case, in addition to these holes, even the fermions with $0<E \le E_+$ are reflected quantum mechanically, and they may also contribute to the thermal radiation.
Indeed, the probability of the reflectance of this process for the fermion with energy $E$ can be calculated as \cite{Moore:1991zv,Giovanazzi:2004zv}
\begin{align}
 P_R(E)= \frac{1}{\exp\left( \frac{2\pi}{\hbar} \sqrt{\frac{m}{\alpha}} E \right)+1}.
\label{PR}
\end{align}
The derivation is shown in appendix \ref{app-QM}.
Again this probability can be interpreted as the Fermi-Dirac distribution at the temperature $T_L$.
Thus the Hawking radiation in the acoustic black hole may be explained as the net effect of the holes and reflected fermions\footnote{Thus the acoustic Hawking radiation is related to the quantum tunneling of the fermions. This result may remind us the derivation of the Hawking radiation through the quantum tunneling proposed by Parikh and Wilczek \cite{Parikh:1999mf}. However they are not directly connected, since the former is the tunneling of the fermions which compose the fluid and the latter is the tunneling of the phonons in the acoustic geometry.}. See figure \ref{Fig-HR}.
Since both the holes and reflected fermions obey the same Fermi-Dirac distribution, the temperature of the bosonic non-critical string case and the acoustic black hole case are the same. 

Now it is natural to interpret the thermal radiation in the bosonic non-critical string case as a variety of the acoustic Hawking radiation.
Although there is no supersonic region in the bosonic non-critical string case, the mechanism of the appearance of the radiation is related to the Hawking radiation in the acoustic black hole.

Finally, in order to confirm that the Hawking radiation in the acoustic black hole can be explained by the quantum mechanics of the fermions, we evaluate the density of the observable $O(\hat{x},\hat{p})$ in the two different ways: the quantum mechanics of the fermions and the second quantized phonon. 
In appendix \ref{app-QM-phonon}, we show that these two derivations provide the same result,
\begin{align*}
 \frac{m}{2\pi \hbar}
 \int_{0}^{\infty} \frac{d E}{e^{\beta_L E}+1}
 \left(
 \frac{O(x,p(E,x))}{|p(E,x)|} 
-  \frac{O(x,p(-E,x))}{|p(-E,x)|} 
\right).
\end{align*}
The first term is the contribution of the reflected fermions and the second one is that of the holes which is equivalent to (\ref{VEV-O-semi}).
(One important point to derive this result  in the phonon calculation is that we need to evaluate the $O(\epsilon^2)$ correction $\rho_2$ and $v_2$ in the $\epsilon$ expansion (\ref{expansion}), since they are the same order to $\langle \psi  \psi  \rangle$ which is the leading contribution of the Hawking effect.)
In this way, we can show that the acoustic Hawking radiation of the phonon is equivalent to the quantum tunneling and reflection of the fermions in the quantum mechanics.

\section{Discussions}
\label{sec-discussion}

We have studied the thermal radiation in the $c=1$ matrix model which is related to the acoustic Hawking radiation in the supersonic fluid.
Remarkably, although the system is not chaotic, the temperature  $T_L$ (\ref{T-mini}) of this radiation saturates the chaos bound (\ref{L-bound}) \cite{Maldacena:2015waa}.
The mechanism of the appearance of the bound\footnote{
	Throughout this paper, we have considered the system at zero temperature.
	We have merely observed that the thermal flux whose temperature saturates the bound (\ref{T-mini}) is induced in the semi-classical regime, and we do not claim that the bound (\ref{T-mini}) is really saturated in our system.
	It would be interesting to understand the relation between the induced thermal flux in our model and the chaos bound \eqref{L-bound} in genuine thermal many body systems.
} in our model is very simple.
The classical motion of the particle in the inverse harmonic potential (\ref{particle-chaos}) receives the quantum correction (\ref{PT}) or (\ref{PR}), and it causes the thermal radiation.
This mechanism would be universal since only the dynamics around the tip of the potential is relevant for the radiation and it does not depend on the details of the potential in the far region.
Thus our result may be applicable to various other systems.
Particularly, applications to the classical chaotic systems would be interesting.
It would be valuable to investigate whether any thermal radiation associated with the classical Lyapunov exponent arises in these systems in the semi-classical regime.
If the mode which classically shows the exponential sensitivity of the initial condition (\ref{chaos}) effectively feels the inverse harmonic potential approximately, thermal radiation might be induced through the quantum effect.

Besides the acoustic black hole may provide a chance to explore the nature of the chaos bound in experiment.
Several interesting phenomena related to chaos may occur near the event horizon of black holes (e.g. \cite{Maldacena:2015waa, Sekino:2008he, Hashimoto:2016dfz,Qaemmaqami:2017jxz}), and some of them may occur even in the acoustic event horizon.
Since acoustic black holes may be realized in laboratories \cite{Lahav:2009wx,Steinhauer:2014dra,Steinhauer:2015saa}, these predictions of the chaotic behaviors might be tested in future.

Finally, understanding the role of the chaos bound \eqref{L-bound} in quantum gravity is a quite important challenge.
Since  the $c=1$ matrix model describes a two dimensional gravity, it is natural to ask whether the bound has any special meaning in this context. 
Naively, one may expect that the bound is related to a black hole in the gravity, but it is hard to consider black holes in the matrix model \cite{Karczmarek:2004bw}.
We leave this problem for future investigations.

\paragraph{Acknowledgements}
The author would like to thank Koji Hashimoto, Pei-Ming Ho, Shoichi Kawamoto, Manas Kulkarni, Gautam Mandal, Yoshinori Matsuo, Joseph Samuel, Tadashi Takayanagi and Asato Tsuchiya for valuable discussions and comments.
The work of T.~M. is supported in part by  Grant-in-Aid for Young Scientists B (No. 15K17643) from JSPS.

\appendix

\section{Quantum mechanics in the inverse harmonic potential}
\label{app-QM}

In this appendix, we review the derivation of the transmittance $P_T(E)$ (\ref{PT}) and the reflectance $P_R(E)$ (\ref{PR}) by solving the Schr\"{o}dinger equation (\ref{H-c=1}). 
We start from the Schr\"{o}dinger equation in the inverse harmonic potential (\ref{H-c=1})
\begin{align}
&E \phi(x)= \hat{H} \phi(x), \quad
\hat{H}=-\frac{\hbar^2 }{2m} \partial_x^2 - \frac{\alpha}{2} x^2
\quad \Rightarrow \quad \phi'' + \frac{m \alpha}{\hbar^2} x^2 \phi + \frac{2m E}{\hbar^2} \phi=0.
\end{align}
By defining the variables
\begin{align}
w=\left(\frac{4m \alpha}{\hbar^2} \right)^{1/4} x, \qquad 
a=- \sqrt{\frac{m}{\alpha}}\frac{E}{\hbar},
\label{w-a}
\end{align}
 the Schr\"{o}dinger equation becomes
\begin{align}
\partial_w^2 \phi + \left(\frac{1}{4}w^2 -a \right)\phi=0.
\end{align}
The two independent solutions of this equation are given by the parabolic cylinder function $E(a,w)$ and its complex conjugate $E^*(a,w)$, where we follow the notation of \cite{Abramowitz}.

To derive the transmittance and the reflectance, we consider the connection between $E(a,w)$ and the WKB wave function
 \begin{align}
\phi_{E \pm}(x)= \frac{A(E) }{\sqrt{p(E,x)}} \exp\left(\pm  \frac{i}{\hbar} \int^x dy~ p(E,y) \right) , \quad
p(E,x)= \sqrt{2mE+m \alpha x^2},
\label{WKB}
\end{align}
where $A(E)$ is a normalization factor which is irrelevant in the following discussion.
As shown in 19.24.1 of \cite{Abramowitz},  $E(a,w)$ oscillates for large $w$ $(\gg \sqrt{|a|})$ as
\begin{align}
& E(a,w)=F e^{i \chi}, \nonumber \\
& \chi \simeq \frac{1}{4}w^2- a \log w + \cdots =\frac{i}{\hbar}\left(
\frac{\sqrt{m\alpha}}{2}x^2+\sqrt{\frac{m}{\alpha}} E \log \left( \left(\frac{4m\alpha}{\hbar^2}\right)^{1/4}x \right)+\cdots \right), \nonumber \\
&  F \simeq \sqrt{\frac{2}{w}}\left(1+ O(w^{-2}) \right) = \frac{\sqrt{\hbar}}{(m \alpha x^2)^{1/4}}+ \cdots .
\label{asymptotic-E}
\end{align}
On the other hand, the momentum $p(E,x)$ in the WKB wave function (\ref{WKB}) behaves 
\begin{align}
p(\epsilon,x) \simeq
\sqrt{m \alpha}x + \sqrt{\frac{m}{\alpha}} \frac{E}{x}  + \cdots,\qquad(x^2 \gg |E|/\alpha).
\end{align}
Thus by comparing this with (\ref{asymptotic-E}), we can read off the connection between $E(a,w)$ and the WKB wave function (\ref{WKB}) as
\begin{align}
E(a,w) \to \frac{\sqrt{\hbar}}{\sqrt{p(E,x)}}  \exp\left(  \frac{i}{\hbar} \int^x dy~ p(E,y) \right), \qquad (w \gg \sqrt{|a|})
\label{E-WKB}.
\end{align}
Hence $E(a,w)$ and $E^*(a,w)$ describe the right ($p>0$) and left ($p<0$) moving wave in $w \gg \sqrt{|a|}$, respectively. 

The final task for obtaining the transmittance and the reflectance is finding the relation between the WKB wave function (\ref{WKB}) in $x \to - \infty$ and in $x \to  \infty$.
For this purpose, the relation 19.18.3 of \cite{Abramowitz} is useful,
\begin{align}
 & \sqrt{1+e^{2\pi a}} E(a,w) = e^{\pi a} E^*(a,w)+i E^*(a,-w), \nonumber \\
\Rightarrow \quad & E(a,w)= i\sqrt{1+e^{2\pi a}} E^*(a,-w)-ie^{\pi a} E(a,-w).
\label{Key}  
\end{align}
Then, for a large negative $w$, we obtain  
\begin{align}
E(a,w) \to i\sqrt{1+e^{2\pi a}} \sqrt{\frac{2}{-w}}e^{-i \left( \frac{w^2}{4}+ \cdots \right) }  -ie^{\pi a} \sqrt{\frac{2}{-w}}e^{i \left( \frac{w^2}{4} + \cdots \right) }  ,
\end{align}
from (\ref{asymptotic-E}).
Note that the momenta carried by the first term and the second term at a large negative $w$ are positive and negative, respectively,  as
\begin{align}
-i \partial_w \left( \sqrt{\frac{2}{-w}}e^{-i \left( \frac{w^2}{4}+ \cdots \right) } \right)
\simeq - \frac{w}{2} \left( \sqrt{\frac{2}{-w}}e^{-i \left( \frac{w^2}{4}+ \cdots \right) } \right), \nonumber \\
-i \partial_w \left( \sqrt{\frac{2}{-w}}e^{i \left( \frac{w^2}{4}+ \cdots \right) } \right)
\simeq  \frac{w}{2} \left( \sqrt{\frac{2}{-w}}e^{i \left( \frac{w^2}{4}+ \cdots \right) } \right) .
\end{align}
Thus the first term represents the in-coming wave towards the inverse harmonic potential and the second term describes the reflected wave.
On the other hand, $E(a,w)$ at large positive $w$ can be regarded as the transmitted wave.
These waves are connected to the WKB wave function (\ref{WKB}) in the asymptotic region as we have seen in (\ref{E-WKB}).

\begin{figure}
  \begin{center}
\includegraphics{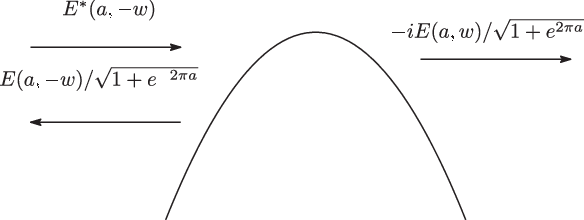}
\caption{The scattering of the in-coming wave $E^*(a,-w)$ through the relation (\ref{scattering}).
$E(a,-w)$  and $E(a,w)$ describe the reflected wave and the transmitted wave, respectively.
}   \label{fig-scattering-2}
  \end{center}
\end{figure}

By normalizing the coefficient of the in-coming wave to 1 in the relation (\ref{Key}), we obtain
\begin{align}
 E^*(a,-w)=& \frac{1}{\sqrt{e^{-2\pi a}+1}}  E(a,-w)   -i \frac{1}{\sqrt{e^{2\pi a}+1}}  E(a,w) \nonumber \\
 =& \frac{1}{\sqrt{\exp\left( \frac{2\pi}{\hbar} \sqrt{\frac{m}{\alpha}} E \right)+1}} E(a,-w)-i  \frac{1}{\sqrt{\exp\left( -\frac{2\pi}{\hbar} \sqrt{\frac{m}{\alpha}} E \right)+1}} E(a,w).
 \label{scattering}
\end{align}
See figure \ref{fig-scattering-2}.
Then, from the coefficients of the reflected wave $E(a,-w)$ and the  transmitted wave $E(a,w)$,
we can read off the reflectance $P_R(E)$ and transmittance $P_T(E)$
\begin{align}
 P_R(E)= \frac{1}{\exp\left( \frac{2\pi}{\hbar} \sqrt{\frac{m}{\alpha}} E \right)+1},  \qquad P_T(E)=\frac{1}{\exp\left( -\frac{2\pi}{\hbar} \sqrt{\frac{m}{\alpha}} E \right)+1},
\label{R-T}
\end{align}
respectively.
Note that $P_R(E)$ and  $P_T(E)$ satisfy $P_T(E) \to 1,~P_R(E) \to 0$ as $E \to + \infty$, and $P_T(E) \to 0,~P_R(E) \to 1$ as $E \to - \infty$.
They agree with the classical motion of the particles.

\section{Comparison of the thermal radiation derived from the quantum mechanics and fluid dynamics}
\label{app-QM-phonon}

As we argued in section \ref{sec-HR-QM}, the temperature derived from the quantum mechanics of the $N$ fermions (\ref{T-mini-2})  and the temperature derived from the Bogoliubov transformation of the phonon (\ref{Hawking-temp})  agree\footnote{The choice of the vacuum in the Bogoliubov transformation would correspond to the choice of the state in the quantum mechanics.
In the Bogoliubov transformation, we implicitly chose the Unruh vacuum to describe the acoustic Hawking radiation.
In the quantum mechanics, this vacuum would correspond to the state in which the fermions flows from the left to right.
This state is not stationary through the quantum effects, and the radiation appears.}.
However you may wonder whether these differently obtained radiation are really equivalent.
In fact, the fermions obey the Fermi-Dirac distribution (\ref{VEV-O-semi}) while the phonons obey the Bose-Einstein distribution. 
Thus it is important to evaluate the radiation quantitatively and confirm whether they agree.
In this appendix, we evaluate the density of the observable $O(\hat{x},\hat{p})$ in the black hole (right) case (\ref{BH-right}) both from the fermions and the phonons, and we will see the agreement.

\subsection{Radiation through the quantum mechanics of the fermions} 
\label{app-HR-QM}

We evaluate  the density of the operator $O(\hat{x},\hat{p})$ in the acoustic black hole case (\ref{BH-right}) by using the quantum mechanics of the fermions.
The derivation is almost same to the derivation of (\ref{VEV-O-semi}) in the bosonic non-critical string case studied in section \ref{sec-c=1}.
As we argued in section \ref{sec-HR-QM}, the difference between the bosonic non-critical string and the acoustic black hole is the existence of the fermions with energy $0<E< E_+$ only. 
Thus what we should do is evaluating the contributions of these fermions to the radiation.

A tiny technical issue for these fermions is the boundary condition of the WKB wave function (\ref{WKB}) at $x=\pm L$.
If we impose $\phi_{E\pm}(L)=\phi_{E\pm}(-L)=0$, ($0<E<E_+$) which we employed for the fermions with $E<0$ in section \ref{sec-c=1},  
the fermions are reflected at $x=\pm L$  and they cannot compose the stationary configuration (\ref{BH-right}) in the classical limit.
In order to avoid this issue, we impose the periodic boundary condition $\phi_{E\pm}(L)=\phi_{E\pm}(-L)$ only for the fermions with positive energy $E$.
Then the configuration (\ref{BH-right}) becomes stationary in the classical limit\footnote{If we impose the periodic boundary condition to the fermions with negative energy, the fermions cannot compose a stationary bosonic non-critical string configuration.
Thus we need to apply the two different boundary conditions depending on the sign of the fermion energy.
We can avoid such a energy dependent boundary condition by employing the periodic potential $V= \alpha (L/\pi)^2 \cos(\pi x/L)$  instead of the inverse harmonic potential and impose the periodic boundary condition $x=x+4L$, and put the fermions appropriately so that the configuration near $x=0$ becomes (\ref{BH-right}).
}.
(The boundary condition at $x=\pm L$ should be irrelevant for the radiation, if $L$ is sufficiently large.)

Then, in the classical limit $(\hbar \to 0)$, the WKB wave function for the  fermion with energy $E>0$ is given by
\begin{align}
\phi_{E \pm}(x)= \frac{A(E) }{\sqrt{p(E,x)}} \exp\left(\pm  \frac{i}{\hbar} \int^x dy~ p(E,y) \right) , \quad
p(E,x)= \sqrt{2mE+m \alpha x^2}.
\label{WKB2}
\end{align}
Here the energy is quantized as
\begin{align}
\int_{-L}^{L} dx\, |p(E,x)| = 2 \pi  \hbar n, \qquad (n=1,2,\cdots),
\label{BS-cond-2}
\end{align}
due to the periodic boundary condition.
The normalization factor $A(E)$ is given by
\begin{align}
|A(E)|^2 = \left( \int_{-L}^{L} \frac{dx}{ |p(E,x)| } \right)^{-1} .
\end{align}
By using this WKB wave function, we evaluate the density of the observable $O(\hat{x},\hat{p})$  in the far region $-L \ll  x \ll  - (\hbar^2/m \alpha)^{1/4}$.
Particularly the leading quantum corrections to this quantity would correspond to the Hawking radiation.
Recall that, in the classical limit ($\hbar \to 0$), the fermions with energy $E>0$ just go through the potential from the left to right, and the left moving fermions appear only through the quantum reflection with the probability $P_R(E)$.
Hence we should evaluate the leading contributions of the left moving fermions to $O(\hat{x},\hat{p})$ to obtain the Hawking radiation.
Through the similar calculation to the $E<0$ case (\ref{O-classical}), such contributions to the density of $O(\hat{x},\hat{p})$  can be obtained as
\begin{align}
 \int_{0}^{E_+} dE \, \rho(E) P_R(E)  |\phi_{E-} (x)|^2
O(x,p(E,x)).
\label{density-O}
\end{align}
Here the energy density $\rho(E)$ is given by
\begin{align}
\rho(E) = \frac{\partial n}{\partial E}= \frac{m}{2\pi \hbar } \int_{-L}^{L} \frac{dx}{ |p(E,x)| }, \qquad (E>0),
\end{align}
through the quantization condition (\ref{BS-cond-2}).
Then the density (\ref{density-O}) can be calculated as
\begin{align}
 \int_{0}^{E_+} dE \, \rho(E) P_R(E)  |\phi_{E-} (x)|^2
O(x,p(E,x))
=
 \frac{m}{2\pi \hbar}
 \int_{0}^{E_+} \frac{d E}{|p(E,x)|} \frac{O(x,p(E,x))}{e^{\beta_L E}+1} .
\end{align}
This is the contribution to the Hawking radiation from the fermions with $0<E\le E_+$.
By assuming that $E_+$ is sufficiently large and can be approximated as $E_+  \approx \infty$, and adding the contribution of the fermions with negative energy (\ref{VEV-O-semi}) with $\mu=0$, we obtain the density of the observable $O(\hat{x},\hat{p})$ induced by the Hawking radiation as
\begin{align}
 \frac{m}{2\pi \hbar}
 \int_{0}^{\infty} \frac{d E}{e^{\beta_L E}+1}
 \left(
 \frac{O(x,p(E,x))}{|p(E,x)|} 
-  \frac{O(x,p(-E,x))}{|p(-E,x)|} 
\right), \qquad -L \ll  x \ll  - \left(\frac{\hbar^2}{m \alpha}\right)^{1/4}.
\label{VEV-O-total}
\end{align}
The first term is the contribution of the reflected fermions and the second term is that of the holes, and both obey the Fermi-Dirac distribution at the temperature $T_L$ (\ref{T-mini}).

More concretely, we take $O(\hat{x},\hat{p})$ as $O(\hat{x},\hat{p})= \hat{x}^k \hat{p}^n$ and evaluate (\ref{VEV-O-total}) as\footnote{We need to take care of the sign of $p=\pm \sqrt{2m(E-V)}$. 
$p$ in $O(x,p)$ at negative $x$ takes a negative value, since the holes and the reflected fermions carry negative momenta. (See figure \ref{Fig-HR}). }
\begin{align}
 \frac{m}{2\pi \hbar} \int_0^\infty   \frac{d E}{e^{ \beta_L E }+1}
 \left(
-x^k p(E,x)^{n-1}+
x^k p(-E,x)^{n-1}
\right)
  =&
- \frac{ (n-1)  \hbar }{48 \pi } \left(m \alpha \right)^{\frac{n-1}{2}} x ^{k+n-3}
 + \cdots .
 \label{vev-QM }
\end{align}
Here we have used $\int_0^\infty \frac{d E \, E}{e^{\beta E}+1}= \frac{ \pi^2}{12 \beta^2} $.
We will compare this result with the quantum field theory calculation of the phonons in the next section.

\subsection{Radiation through quantum field theory of  phonons} 
\label{app-HR-fluid}

We evaluate the observable $O(\hat{x},\hat{p})= \hat{x}^k \hat{p}^n$ in the acoustic black hole through the quantum field theory computation.
We introduce the Wigner phase space density $u(x,p)$ which classically satisfies
\begin{align}
u(x,p) & =
\begin{cases}
 1 & \quad \text{If the point }(x,p)\text{ in the phase space is occupied by a fermion.}  \\
  0 &  \quad \text{If the point }(x,p)\text{ in the phase space is unoccupied.}
\end{cases}
\end{align}
By using this function, we can evaluate the density of the observable $O(\hat{x},\hat{p})$ at position $x$ in the classical limit as \cite{Dhar:1992rs,Dhar:1992hr,Mandal:2013id}
 \begin{align}
 \frac{1}{2\pi \hbar} \int_{-\infty}^{\infty} d p \,  u(p,x) O(x,p)
  = \frac{1}{2\pi \hbar} \int_{P_{0-}(x)}^{P_{0+}(x)} d p \, x^k p^{n}
 = \frac{1}{2\pi \hbar} \frac{x^k}{n+1} \left(  P_{0+}^{n+1}(x)- P_{0-}^{n+1}(x) \right)
.
\label{O-classical-fluid}
\end{align}
Here $P_{0\pm}(x)$ are the acoustic black hole background given in (\ref{BH-right}).

Now we consider the quantum corrections to this equation.
In quantum field theory,  $P_{0\pm}^{n+1}$ should be replaced by $\langle P_{\pm}^{n+1} \rangle$, where the expectation value is taken by the vacuum for the phonon discussed in section \ref{sec-HR} which satisfies
\begin{align}
 \langle P_- \rangle \to P_{0-}, \qquad \hbar \to 0, \quad \text{and} \quad  \langle P_+ \rangle = P_{0+}.
 \label{P-classical} 
\end{align}
The contributions of the Hawking radiation would be the leading quantum corrections to the classical result (\ref{O-classical-fluid}), and we evaluate them as
\begin{align}
& \frac{1}{2\pi \hbar} \frac{x^k}{n+1} \left( \langle  P_{+}^{n+1}(x) \rangle - \langle P_{-}^{n+1}(x) \rangle \right)
-\frac{1}{2\pi \hbar} \frac{x^k}{n+1} \left(  P_{0+}^{n+1}(x)- P_{0-}^{n+1}(x) \right)
\nonumber  \\
 =& 
- \frac{x^k}{2\pi \hbar}\left(  \epsilon P_{0-}^{n} \langle P_{1-} \rangle  + \epsilon^2 \left( \frac{n}{2} P_{0-}^{n-1} \langle P_{1-}^2 \rangle  +   P_{0-}^{n} \langle P_{2-} \rangle  \right) + \cdots \right)   .
 \label{vev-hydro}
\end{align}
Here we have kept the terms up to $O(\epsilon^2)$ in the $\epsilon$ expansion (\ref{expand-P}).
From now, we take $\epsilon=1$, since $\hbar$ will play the role of a natural expansion parameter in the $\hbar$ expansion around $P_{0\pm}$ (\ref{P-classical}). 
By using the relation (\ref{sol-P}), we obtain the expectation values of $P_{1-}$ as
\begin{align}
\langle P_{1-} \rangle = 
\frac{2m^2}{P_{0-} } 
\langle  \partial_{z_-} \psi_-   \rangle=0
, \qquad 
 \langle P_{1-}^2 \rangle = \frac{4m^4}{P_{0-}^2 } 
\langle  :\partial_{z_-} \psi_-  \partial_{z_-} \psi_- :  \rangle .
\end{align}
Hence we need to evaluate $\langle  :\partial_{z_-} \psi_-  \partial_{z_-} \psi_- :  \rangle
$.
Although we can calculate it by using the Bogoliubov transformation,
here, instead, we take a shortcut by using the relation to the energy momentum tensor\footnote{Here the coefficient $m^2/ \hbar \pi$ in (\ref{EM-psi}) can be read off from the Hamiltonian of the fluid variables \cite{Ginsparg:1993is,Mandal:2013id,Polchinski:1991uq}
\begin{align}
H&= \int dx\   \rho(x)  \left( \frac12 m  v(x)^2 +  \frac{\hbar^2 \pi^2}{6m}  \rho(x)^2 +V(x)\right).
\end{align}
By substituting the expansion (\ref{expansion}) to this Hamiltonian, we can derive the action for $\psi$ as
\begin{align}
S[\psi]= \int dtdx \sqrt{-g}  \frac{m^2}{2\pi \hbar} g^{\mu\nu} \partial_\mu \psi \partial_\nu \psi,
\end{align}
and we obtain the energy momentum tensor (\ref{EM-psi}).
}
\begin{align}
T_{z_- z_-} = \frac{m^2}{ \hbar \pi}
 :\partial_{z_-} \psi_-  \partial_{z_-} \psi_- : .
\label{EM-psi}
 \end{align}
 Then, from (\ref{EM-tensor}), we obtain
\begin{align}
 \langle P_{1-}^2 \rangle = \frac{4m^4}{P_{0-}^2 } \frac{ \hbar \pi}{m^2}
\langle T_{z_-z_-} \rangle =
\frac{4m^2 \hbar \pi}{P_{0-}^2 } 
 \frac{\hbar}{48\pi} \frac{\alpha}{m}.
\end{align}

Next we calculate $\langle P_{2-} \rangle$.
By applying the $\epsilon$ expansion (\ref{expand-P}) to the equation for $P_-$ (\ref{eom-P}), we obtain \begin{align}
&0=\partial_t P_{2-} + \frac{1}{m} \partial_x \left(P_{0-} P_{2-} + \frac{1}{2}P_{1-}^2  \right), \nonumber \\
&\Rightarrow
\left( \partial_t  + \frac{P_{0-}}{m} \partial_x \right) \left(P_{0-} P_{2-} \right)=- \frac{1}{2}
\left( \partial_t  + \frac{P_{0-}}{m} \partial_x \right) \left(P_{1-}^2  \right)
+ \frac{1}{2} \partial_t \left(P_{1-}^2  \right),
\end{align}
where we have used $\partial_t P_{0-}=0$.
Since $\partial_t \langle P_{1-}^2 \rangle =0$, by integrating this equation, we obtain
\begin{align}
 \langle P_{2-} \rangle = - \frac{1}{2P_{0-}}  \langle P_{1-}^2 \rangle .
\end{align}
By substituting these results to (\ref{vev-hydro}), we obtain
\begin{align}
&-  \frac{x^k}{2\pi \hbar}\left(  P_{0-}^{n} \langle P_{1-} \rangle  +  \left( \frac{n}{2} P_{0-}^{n-1} \langle P_{1-}^2 \rangle  +   P_{0-}^{n} \langle P_{2-} \rangle  \right) + \cdots \right)   \nonumber \\
=&- \frac{x^k}{2\pi \hbar} \left( \frac{n-1}{2} P_{0-}^{n-1} \langle P_{1-}^2 \rangle   + \cdots  \right)
= -\frac{ (n-1)  \hbar }{48 \pi } \left(m \alpha \right)^{\frac{n-1}{2}} x ^{k+n-3}
 + \cdots ,
\end{align}
for the large negative $x$.
Here we have used (\ref{BH-right}) for $P_{0-}$.
This agrees with the result (\ref{vev-QM }) obtained from the quantum mechanics.
Therefore the Hawking radiation obtained from the quantum mechanics and quantum field theory are consistent.

{\normalsize \bibliographystyle{unsrt} \bibliography{AHR} }

\end{document}